\documentclass[journal]{IEEEtran}

\usepackage{cite}

\ifCLASSINFOpdf
  \usepackage[pdftex]{graphicx}
 
\else
  \usepackage[dvips]{graphicx}
\fi

\usepackage{amsmath}
\ifCLASSOPTIONcompsoc
  \usepackage[caption=false,font=normalsize,labelfont=sf,textfont=sf]{subfig}
\else
  \usepackage[caption=false,font=footnotesize]{subfig}
\fi

\usepackage{listings}
\usepackage{cleveref}
\crefname{equation}{}{}
\Crefname{equation}{Equation}{Equations}
\crefname{figure}{Fig.}{Figs.}
\Crefname{figure}{Fig.}{Figs.}

\newcommand{\bigO}[1]{\mathcal{O}(#1)}

\usepackage{tikz}
\usepackage{textcomp}

\begin{document}

\title{A Fast Algorithm for\\Calculation of Th\^eo1}

\author{Ben~Lewis
\thanks{Department of Physics, University of Strathclyde, UK, email: b.lewis@strath.ac.uk}
\thanks{Manuscript received 19 May 2020}}

\markboth{IEEE Transactions on Ultrasonics, Ferroelectrics and Frequency Control}{}%
%

\newcommand\copyrighttext{%
  \footnotesize \copyright 2020 IEEE. Personal use of this material is permitted.
  Permission from IEEE must be obtained for all other uses, in any current or future
  media, including reprinting/republishing this material for advertising or promotional
  purposes, creating new collective works, for resale or redistribution to servers or
  lists, or reuse of any copyrighted component of this work in other works.
  DOI: 0.1109/TUFFC.2020.2996313}
\newcommand\copyrightnotice{%
\begin{tikzpicture}[remember picture,overlay]
\node[anchor=south,yshift=10pt] at (current page.south) {\fbox{\parbox{\dimexpr\textwidth-\fboxsep-\fboxrule\relax}{\copyrighttext}}};
\end{tikzpicture}%
}

\maketitle 
\copyrightnotice
\begin{abstract}
Th\^eo1 is a frequency stability statistic which is similar to the Allan variance but can provide stability estimates at longer averaging factors and with higher confidence. However, the calculation of Th\^eo1 is significantly slower than the Allan variance, particularly for large data sets, due to a worse computational complexity. A faster algorithm for calculating the `all-$\tau$' version of Th\^eo1 is developed by identifying certain repeated sums and removing them with a recurrence relation. The new algorithm has a reduced computational complexity, equal to that of the Allan variance. Computation time is reduced by orders of magnitude for many datasets. The new, faster algorithm does introduce an error due to accumulated floating point errors in very large datasets. The error can be compensated for by increasing the numerical precision used at critical steps. The new algorithm can also be used to increase the speed of Th\^eoBr and Th\^eoH which are more sophisticated statistics derived from Th\^eo1.
\end{abstract}


%
\IEEEpeerreviewmaketitle

\section{Introduction}
%
%
%
%
\IEEEPARstart{W}{hen} characterising a frequency source, it is necessary to determine the amount and types of noise which determine the frequency stability on different timescales. There are many statistics used for this purpose, one of which is the `theoretical variance \#1' (Th\^eo1)\cite{howe_very_2003}. Compared to the more commonly used Allan variance, Th\^eo1 has increased confidence at long averaging times, and can be used to estimate stability up to 50\% longer averaging times. Th\^eo1 is also better able to identify which type of `power-law' noise is present\cite{mcgee_howe_2007,schwindt_jau_2015}. These properties have allowed Th\^eo1 to be used for long-running experiments where datasets cannot easily be extended\cite{Parker_2005}. However, Th\^eo1 is slow to compute for large datasets, as its computational complexity is $\bigO{N^3}$ for a dataset of $N$ measurements. Similar statistics with $\bigO{N^3}$ complexity have been reduced to $\bigO{N^2}$ complexity by the use of an appropriate algorithm\cite{li_peng_2000}.

For a series of $N$ time deviation points $x_i$, each separated by an interval $\tau_0$, Th\^eo1 can be defined\cite{howe_very_2003} as
\begin{equation}
    \label{eqn:Theo_defn}
    \text{Th\^eo1}(\tau=1.5k\tau_{0},N) = \frac{T_k}{3(N-2k)(k\tau_0)^{2}}
\end{equation}
where $0<k\leq(N-1)/2$, the averaging time is $\tau$ and
\begin{multline}
    \label{eqn:T_defn_2}
     T_k = \sum_{i=0}^{N-2k-1}\sum_{\delta=0}^{k-1}\\
     \frac{1}{(k-\delta)}[(x_{i}-x_{i-\delta+k})+(x_{i+2k}-x_{i+\delta+k})]^2
\end{multline}

A naive implementation of this definition of Th\^eo1 will have a complexity of $\bigO{N^3}$ because there are $\approx N/2$ values of $k$ for which to calculate $T_k$, each taking $\bigO{N^2}$ operations due to the nested sum in \cref{eqn:T_defn_2}.
This can make computation prohibitively expensive for extremely large datasets or in applications requiring low latency, such as measuring the dynamic stability of an oscillator with a high data rate\cite{howe_limited_2011}.

In some cases it is not necessary to calculate Th\^eo1 for every value of $k$, sometimes called an `all-$\tau$' calculation,  and it may be sufficient to only use $k$ equal to powers of two. However, the more sophisticated statistics Th\^eo{Br} and Th\^eo{H}\cite{howe_theoh_2006}, which attempt to correct for bias in Th\^eo1 require the calculation of Th\^eo1 for all $k$ as a first step. There is a technique called `fast Theo{Br}' \cite{taylor_fast_2010} which increases the speed of this calculation by averaging points within the initial dataset to reduce its size. However, for a fixed amount of averaging, the speed increase is only a constant factor and does not change the $\bigO{N^3}$ complexity.

\IEEEpubidadjcol

\section{Algorithm}
One way to produce a faster algorithm for Th\^eo1 is to find a recurrence relation between parts of the outer sum, which allows calculation of one term from the next without performing the full inner sum.
This is made difficult by the term $1/(k-\delta)$ which forces a different coefficient before each terms as $\delta$ is incremented in the inner sum.
However, the definition of $T_k$ can be rearranged to move this awkward term outside the inner sum by swapping the order of the sums and using the substitution $v = k-\delta$, so that
\begin{equation}
    \label{eqn:T_defn}
     T_{k}   = \sum_{v=1}^{k} \frac{1}{v}A_{k,v}\\
\end{equation}
where $A_{k,v}$ is defined by
\begin{align}
    A_{k,v} =& \sum_{i=0}^{N-2k-1}  (x_i-x_{i+v}+x_{i+2k}-x_{i+2k-v})^2 \label{eqn:A_defn_1}\\ 
            =& \sum_{i=0}^{N-2k-1} \label{eqn:A_defn_2}
            \begin{aligned}[t]
            &(x_i^2+x_{i+v}^2+x_{i+2k}^2+x_{i+2k-v}^2\\
            &+2x_ix_{i+2k}+2x_{i+v}x_{i+2k-v}\\
            &-2x_ix_{i+v}-2x_ix_{i+2k-v}\\
            &-2x_{i+v}x_{i+2k}-2x_{i+2k}x_{i+2k-v})\,.
            \end{aligned}
\end{align}
Some of the expanded terms in \cref{eqn:A_defn_2} have similar forms, and can be expressed in terms of new summations $C^{(n)}$, defined as

\begin{align}
    C^{(1)}_j =& \sum_{i=0}^{j}x_i^2\label{eqn:C1_defn}\\
    C^{(2)}_j =& \sum_{i=0}^{N-j-1}x_i x_{i+j}\label{eqn:C2_defn}\\
    C^{(3)}_{k,j} =& \sum_{i=k}^{N-k-1}x_{i-j}x_{i+j}\label{eqn:C3_defn}\\
    C^{(4)}_{k,j} =& \sum_{i=0}^{N-2k-1}x_{i}x_{i+j}+x_{i+2k}x_{i+2k-j}\,.\label{eqn:C4_defn}
\end{align}
It can then be shown by substitution that
\begin{equation}
    \label{eqn:A_defn_3}
    A_{k,v} =\,\begin{aligned}[t] &C^{(1)}_{N-2k-1}+(C^{(1)}_{N-2k-1+v}-C^{(1)}_{v-1})\\
    &+(C^{(1)}_{N-1}-C^{(1)}_{2k-1})+(C^{(1)}_{N-v-1}-C^{(1)}_{2k-v-1})\\
    &+2(C^{(2)}_{2k}+C^{(3)}_{k,k-v}-C^{(4)}_{k,v}-C^{(4)}_{k,2k-v})\,.\end{aligned}
\end{equation}

The calculation of $T_k$ from the $C^{(n)}$ can be completed in $\bigO{N^2}$, so if the $C^{(n)}$ could all be calculated in $\bigO{N^2}$ then this would reduce the overall complexity of Th\^eo1 to $\bigO{N^2}$. For $C^{(1,2)}$ the definition is already $\leq\bigO{N^2}$, but it can also be achieved for $C^{(3,4)}$ by using a recurrence relation between consecutive terms to avoid the full sum in \cref{eqn:C3_defn,eqn:C4_defn}:
\begin{align}
    C^{(3)}_{k,j} =&\label{eqn:C3_recur}
    \begin{aligned}[t]
        &C^{(3)}_{k-1,j} - x_{k-1-j}x_{k-1+j}\\
        &- x_{N-k-j}x_{N-k+j}
    \end{aligned}
    & ,j <k\\[10pt]
    C^{(4)}_{k,j} =&\label{eqn:C4_recur}
    \begin{aligned}[t]
        &C^{(4)}_{k-1,j} - x_{2k-2-j}x_{2k-2}\\
        &- x_{2k-1-j}x_{2k-1} - x_{N-2k}x_{N-2k+j}\\
        &- x_{N-2k+1}x_{N-2k+1-j}
    \end{aligned}
    & ,j <2k-1
\end{align}
This allows almost all values of $C^{(3,4)}$ to be calculated in a recursive manner, the remaining values are
\begin{align}
    &C^{(3)}_{k,k}      &=\,& C^{(2)}_{2k}&\label{eqn:C3_rem}\\
    &C^{(4)}_{k,2k-1}   &=\,& 2C^{(2)}_{2k-1}-x_0x_{2k-1}-x_{N-2k}x_{N-1}&\label{eqn:C4_rem_1}\\
    &C^{(4)}_{k,2k}     &=\,& 2C^{(2)}_{2k}&\label{eqn:C4_rem_2}
\end{align}
and so $C^{(3,4)}$ can be calculated in the required $\bigO{N^2}$. Because the recurrence relations are for an incremented $k$ value, the technique can only be used when calculating $T_k$ for all values of $k$.

\begin{figure}[t]
    \centering
    \includegraphics{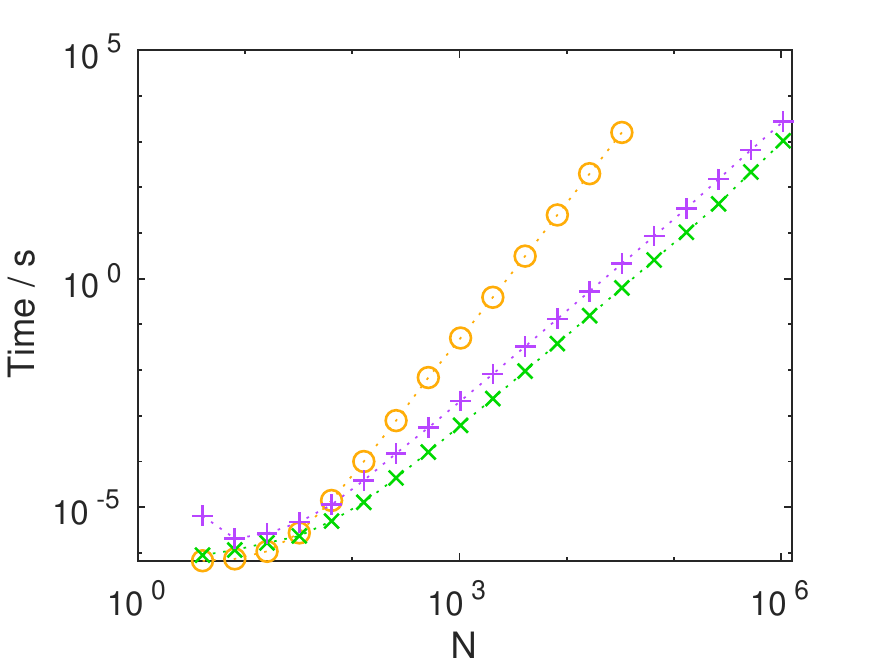}
    \caption{The time taken by different methods of calculating Th\^eo1. Orange circles show the naive method. Green 'x' marks show the new algorithm using a standard double precision floating point datatype. Purple `+' marks show the new algorithm using the int-128 datatype.}
    \label{fig:speed}
\end{figure}

In order to calculate $T_k$ it is sufficient to calculate
 \begin{align}
    C^{(1)}_j,&\quad 0\leq j \leq  N\\
    C^{(2)}_j,&\quad j=2k,2k-1\\
    C^{(3)}_{k,j},&\quad 0\leq j \leq  k\\
    C^{(4)}_{k,j},&\quad 0\leq j \leq  2k
\end{align}
so only these values need to be held in memory. The recursion relations \cref{eqn:C3_recur,eqn:C4_recur} can then be used to update $C^{(3,4)}_{k,j}$ to $C^{(3,4)}_{k+1,j}$ in place, allowing calculation of $T_{k+1}$. The memory requirement is only $\bigO{N}$. Specifically,  it requires memory for $4N$ double precision values: $N$ values for each of the input array $x$ and $C^{(1,4)}$, and $N/2$ values for each of $C^{(3)}$ and the output array. The naive algorithm requires storage of $3N/2$ values so this is a significant increase but is still only 32 MB for a dataset with $N=10^6$.

In order to calculate Th\^eo1, the algorithm can proceed as follows:
\begin{enumerate}
    \item Calculate $C^{(1)}$ using \cref{eqn:C1_defn}.
    \item For each value of $k$ from $0$ to $\lfloor{(N-1)/2}\rfloor\,:$
    \begin{enumerate}
        \item Calculate required values of $C^{(2)}$ using \cref{eqn:C2_defn}.  
        \item Add new values to $C^{(3,4)}$ using \cref{eqn:C3_rem,eqn:C4_rem_1,eqn:C4_rem_2}.
        \item Update $C^{(3,4)}$ using \cref{eqn:C3_recur,eqn:C4_recur}.
        \item Calculate $A_{k,v}$ from the $C^{(n)}$ using \cref{eqn:A_defn_3}.
        \item Calculate $T_k$ from $A_{k,v}$ using \cref{eqn:T_defn}.
    \end{enumerate}
\end{enumerate}

An example implementation of the algorithm in C++ can be found in \cref{appendix:implementation}.

\section{Accuracy}

Whilst the new algorithm for Th\^eo1 is faster than the naive approach, it has more opportunities for floating point errors to accumulate. \Cref{eqn:A_defn_3} shows that terms of similar magnitude are subtracted from each other, allowing catastrophic cancellation to occur and leading to a loss of precision. 
The size of each term in \cref{eqn:A_defn_3} is~$\leq \sum x^2$ and the size of the total is $\sim T_k$, so the fractional error might be expected to scale as~$\sim \langle x^2\rangle /T_k $.
Th\^eo1 is insensitive to any offset or linear change in $x$, so these components can be removed in order to reduce~$\langle x^2 \rangle$ without ill-effect.
This prevents a significant drop in precision that could be caused by a constant frequency or phase offset
This change alone is sufficient to prevent appreciable errors in most practical situations. However, in some cases with large datasets and where the long-term clock stability is dominated by frequency drift the errors could grow large enough to be significant.

\begin{figure}[t]
    \centering
    \includegraphics{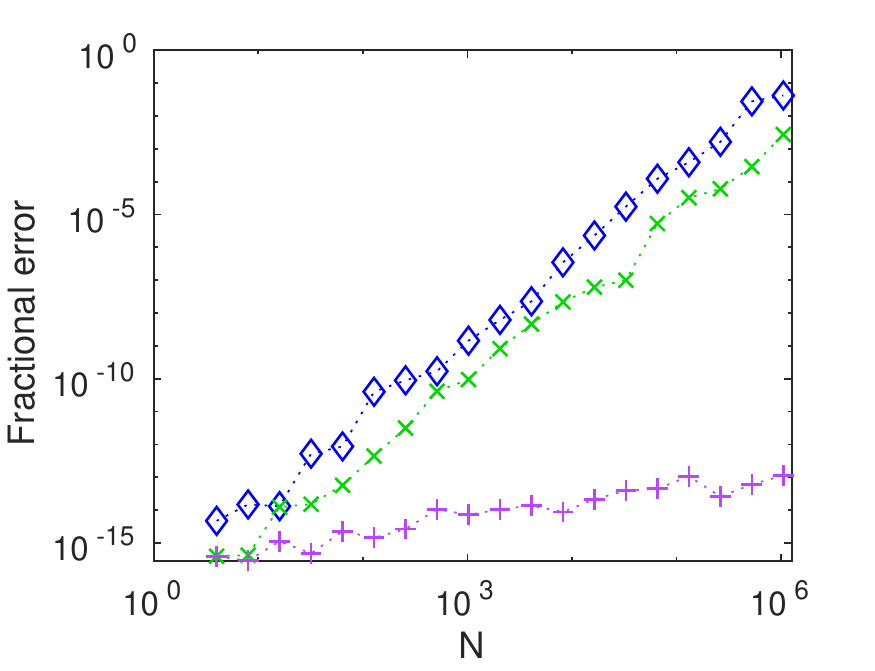}
    \caption{The maximum fractional error introduced by different methods of calculating Th\^eo1, measured against the naive method. Green `x' marks (blue diamonds) indicate the standard double precision method with (without) any linear component removed. Purple `+' marks used the int-128 datatype for increased accuracy. Results depend strongly on the type of noise simulated, for this test white frequency noise with a linear frequency drift was used.}
    \label{fig:accuracy}
\end{figure}

In order to fully mitigate the floating point error it is necessary to use additional bits of precision.
Whilst 128-bit floating point types are available in some environments, they are not generally supported in hardware and are therefore very slow to use.
In contrast, many 64-bit CPUs have hardware support for multiplying two 64-bit integers into a 128-bit integer.
By re-scaling the data and converting it to a 64-bit integer representation, this CPU instruction can be used to calculate Th\^eo1 quickly without floating point errors.
Terms $\propto x$ are stored as 64-bit integers, but terms $\propto x^2$ are stored as 128-bit integers.
In C or C++ the GCC \lstinline{__int128} datatype may be used with multiplications implemented as
\begin{lstlisting}[language=C++]
int64_t x1,x2;
__int128 result = (__int128) x1 * x2;
\end{lstlisting}
The use of a larger data types does cause a speed reduction of approximately 70\%, possibly due to additional memory overhead.
The conversion between datatypes is $\bigO{N}$ and takes negligible time in most cases. 

The speed and fractional floating point error for the different methods are shown in \cref{fig:speed,fig:accuracy} respectively.
The fractional error was measured by comparing the value of Th\^eo1 as calculated with the new algorithm to that calculated with the naive algorithm.
Due to the slow speed of the naive algorithm, only points spaced at powers of two were compared, and the maximum of these errors was taken.
\Cref{fig:accuracy} should be taken as indicative only as the details vary significantly depending on the noise type of the simulated data, although  the `int-128' method had negligible error in all cases tested. To exaggerate the errors seen, a white frequency noise with added linear frequency drift was simulated. The linear drift was chosen such that the frequency stability at the longest and shortest averaging factors was approximately equal. A linear frequency drift is particularly difficult for the simpler error reduction method (removing any linear component to $x$) to deal with, as the dominant $x$ component is quadratic. Despite this, 1 to 2 orders of magnitude improvement was seen.

\section{Conclusion}
Manipulating the definition of Th\^eo1 has led to an algorithm that calculates the `all-$\tau$' version with a reduced computational complexity of $\bigO{N^2}$. Although the new algorithm initially lead to loss of precision, this is reduced by removing any linear component in the dataset. In the cases where the error is still significant, it is made negligible by using an int-128 datatype, at the cost of a $\approx 70\%$ slowdown.

This algorithm makes the speed of Th\^eo1 (and Th\^eo{Br} or Th\^eo{H})  calculations similar to other time-domain stability statistics such as the Allan variance. The new algorithm can more easily be used in low-latency applications such as characterising oscillators using a high sample rate, and commercial testing of oscillators and other sensors. It also makes the use of dynamic Th\^eo{} statistics easier, with opportunities for calculating statistics over multiple timescales simultaneously. Further work could incorporate this algorithm into a dynamic Th\^eo algorithm and be used to identify changes to clock stability in real-time.
 
 All data and code supporting this publication are openly available from the University of Strathclyde KnowledgeBase at https://doi.org/10.15129/4403d30e-4257-4a4a-817f-f459e7465011.
 
\appendices
\crefalias{section}{appendix}
\section{Example implementation}
\label{appendix:implementation}
\begin{lstlisting}[language=C++,lastline=26]
/*========================================
* Calculates Theo1 of a given dataset. 
* The dataset is assumed to be phase
* data, taken at unit time increments.
* Any linear component to the dataset is
* removed as a first step. Should be 
* compiled with -O3 -ffast-math.
*
* X = input dataset
* N = number of elements in X
* T = Theo1 output
*
* Ben Lewis, University of Strathclyde 
*======================================*/

/* Remove any linear part of the dataset*/
void removeLinear(double* X,int N){
    double midpoint = ((double) (N-1))/2;
    long double sum1 = 0;
    long double sum2 = 0;
    for(int i = 0; i < N; i++){
        sum1 += X[i];
        sum2 += X[i]*(i-midpoint);
    }
    double a = sum1/N;
    double b = sum2/N*12/((double)N*N-1);
\end{lstlisting}
\newpage
\begin{lstlisting}[language=C++,firstline=27]
    for(int i = 0; i < N; i++){
        X[i] -= a + b*(i-midpoint);
    }
}

/* The computational routine */
void Theo1(double *X, double *T, int N){
// Initialise arrays
int k_max = (N-1)/2;
double* C1 = new double[N];
double* C3 = new double[k_max+1];
double* C4 = new double[k_max*2];
// Preprocess by removing linear part
removeLinear(X,N);
// Calculate C1
double s=0;
for(int i = 0; i < N; i++){
  s += (X[i]*X[i]);
  C1[i] = s;
}
// Main loop
C3[0] = C1[N-1];
for(int k=1; k<=k_max; k++){
  //Calculate C2 values
  double C2_2k = 0;
  double C2_2k_1 = 0;
  for(int j = 0; j <= N-2*k-1; j++){
    C2_2k += (X[j]*X[j+2*k]);
    C2_2k_1 += (X[j]*X[j+2*k-1]);
  }
  C2_2k_1 += (X[N-2*k]*X[N-1]);
  // Update C3, C4 in place
  for(int v=0; v < k; v++){
    C3[v] -= (X[k-1-v]*X[k-1+v])
           + (X[N-k+v]*X[N-k-v]);
  }
  for(int v = 1;v<=2*k-2;v++){
    C4[v-1] -= (X[2*k-1-v]*X[2*k-1])
             + (X[2*k-2-v]*X[2*k-2])
             + (X[N-2*k]*X[N-2*k+v])
             + (X[N-2*k+1]*X[N-2*k+1+v]);   
  }
  C3[k] = C2_2k;
  C4[2*k-2] = 2*C2_2k_1 - (X[0]*X[2*k-1])
            - (X[N-2*k]*X[N-1]);
  C4[2*k-1] = 2*C2_2k;
  //Calculate un-normalised T_k from C1-C4
  double T_k = 0;
  double A0 = C1[N-1] - C1[2*k-1]
            + C1[N-2*k-1] + 2*C2_2k;
  for(int v = 1;v<=k;v++){
    double A1 = A0 - C1[v-1]
              + C1[N-1-v] - C1[2*k-v-1]
              + C1[N-1-2*k+v];
    double A2 = C3[k-v] - C4[v-1]
              - C4[2*k-v-1];
    T_k += (A1+2*A2)/v;
  }
  // Apply normalisation to get Theo1
  T[k-1] = (T_k/(3*(double)(N-2*k)*k*k));
}
//Release memory
delete[] C1;
delete[] C3;
delete[] C4;
}
\end{lstlisting}
\section*{Acknowledgment}

I would like to thank David Howe and Magnus Danielson for their discussions regarding this research.




\bibliographystyle{IEEEtran}
\bibliography{IEEEabrv,references.bib}

%

%

%

\begin{IEEEbiography}[{\includegraphics[width=1in,height=1.25in,clip,keepaspectratio]{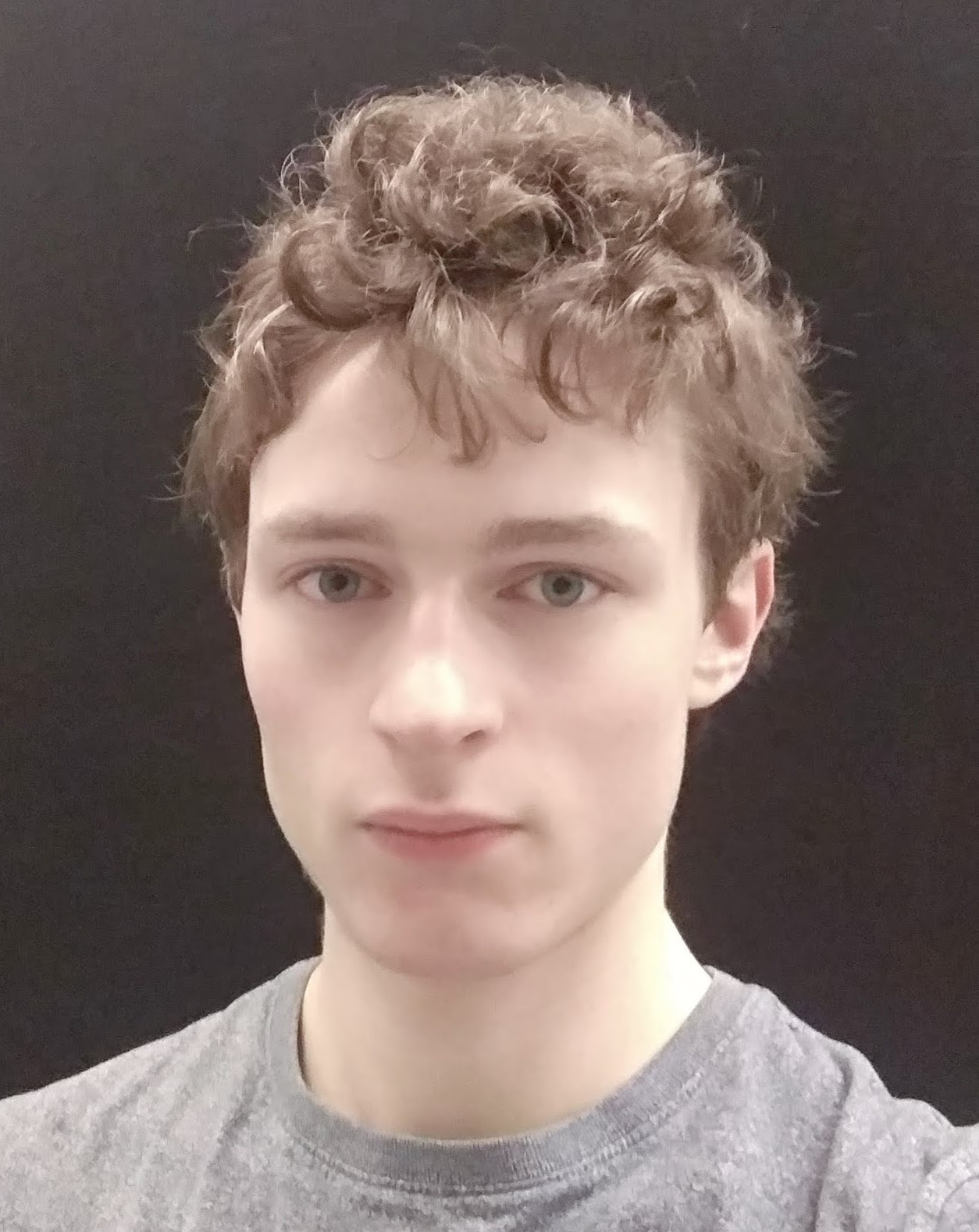}}]{Ben~Lewis} 
received the MSci and BA degrees in Natural Sciences from the University of Cambridge, Cambridge, UK in 2018. He has been a PhD student in the Experimental Quantum Optics and Photonics group at the University of Strathclyde, Glasgow, UK since 2018. His research is focused on compact cold-atom clocks.
\end{IEEEbiography}
\vfill



\end{document}